\documentclass[prl,twocolumn,showpacs,preprintnumbers,amsmath,amssymb]{revtex4}

\usepackage{bm}
\usepackage{amsmath,amssymb,amsthm,times,graphics,graphicx,bm,xcolor}
\usepackage[normalem]{ulem}

\begin{document}

\title{Sagnac Interferometer and the Quantum Nature of Gravity}

\author{Chiara Marletto and Vlatko Vedral}
\affiliation{Clarendon Laboratory, University of Oxford, Parks Road, Oxford OX1 3PU, United Kingdom and\\Centre for Quantum Technologies, National University of Singapore, 3 Science Drive 2, Singapore 117543 and\\
Department of Physics, National University of Singapore, 2 Science Drive 3, Singapore 117542 and\\Fondazione ISI, Torino}

\date{\today}

\begin{abstract}
We use a quantum variant of the Sagnac interferometer to argue for the quantum nature of gravity as well as to formulate a quantum version of the equivalence principle. 
We first present an original derivation of the phase acquired in the conventional Sagnac matter-wave interferometer, within the Hamiltonian formalism. Then we modify the interferometer in two crucial respects. The
interfering matter wave is interfered along two different distances from the centre and the interferometer is prepared in a superposition of two different angular velocities. We argue that
if the radial and angular degrees of freedom of the matter wave become entangled through this experiment, then, via the equivalence principle, the gravitational field must be non-classical. 
\end{abstract}


\maketitle                           

Einstein's happiest thought of his life was that a falling observer is actually an inertial one, i.e. that if we were placed in a falling elevator we would not experience gravity. This led Einstein to formulate the equivalence principle according to which
acceleration is locally indistinguishable from gravity. Once that is established, it is clear that gravity can no longer be a scalar, as it happens to be in Newtonian physics. One number associated to every point in
space (and every instant of time) is insufficient to capture all the relevant aspects of gravity since an accelerated observer, standing
somewhere away from the centre on a rotating disk, experiences not just the centrifugal force (pushing her away from the centre), but also the Coriolis force (therefore one needs at least two numbers associated
with every point on the disk). When this argument is extended to generally accelerated frames, it turns out one needs a two component tensor, known as the metric tensor $g_{\mu\nu}$, to describe fully
the gravitational field. 

We have previously used two massive particle interferometers to argue that if the two particles only couple through the gravitational field, any entanglement generated between them would be evidence
of non-classicality of the gravitational field, \cite{SOUG, MAVE}. Now we will show that, following the equivalence principle, we can make the same argument with a single particle matter wave Sagnac interferometer, but with
a twist that the interferometer has to be prepared in a superposition of two different angular frequencies. This argument is interesting for two reasons. The first one is that unlike in our previous proposal, here there is only
one interfering mass. The superposition of geometries is created by the disk spinning in a superposition of two angular frequencies. The second point is that the entanglement created is between different degrees of freedom
of one and the same particle. However, if our argument is correct, it must also be mediated by geometry, which therefore cannot be classical (if indeed it results in an entangled state). Hereinafter, we will take the equivalence principle in the presence of superposed spacetime configurations to mean that in each branch of the superposition gravity and acceleration are locally indistinguishable. (There is currently a wide debate about how to formulate the equivalence principle in quantum theory \cite{BEI, HARDY, BRU1, BRU2}, but here we assume this particular version of the principle and explore its consequences.)

Let us first review the standard Sagnac matter wave interferometer, \cite{SAGNAC}, but emphasising in the derivation of the interference the analogous nature of acceleration and gravity. The metric on a rotating disk is
given by:
\[
g_{\mu\nu} =   \begin{bmatrix}
     -1 +\frac{\Omega^2 r^2}{c^2} & 0 & \frac{\Omega r^2}{c} & 0 \\
     0 & 1 & 0 & 0 \\
     \frac{\Omega r^2}{c} & 0 & r^2 & 0 \\
     0 & 0 & 0 & 1 
   \end{bmatrix}
\] 
where $\Omega$ is the angular frequency of the disk, $r$ is the coordinate labelling the distance from the centre and $c$ is the speed of light. We can write the metric in the linear regime away from the
flat Minkowski metric as $g_{\mu\nu} = \delta_{\mu\nu}+h_{\mu\nu}$, where $h_{\mu\nu}$ is the deviation from the Minkowski metric $\delta_{\mu\nu}$. In this notation, the relevant linear element will be $h_{00} =  \frac{\Omega^2 r^2}{c^2}$. We will now start the particle at some angle,
a distance $r$ away, and coherently split it to take two different paths (in a superposition) along the positive and negative directions (i.e. clockwise and anti-clockwise). 

The phase difference obtained along two paths that meet after competing a full circle is given by 
\begin{equation}
\Delta \phi = \frac{T_{00}h_{00}}{\hbar} t = \frac{mc^2}{\hbar} \times \frac{\Omega^2 r^2}{c^2} \times \frac{2\pi}{\Omega} = \frac{2m}{\hbar} \times \Omega \times A 
\end{equation}
where $A=\pi r^2$ is the area enclosed by the interfering matter wave. Here we have simply used the fact that the evolution of the system is given by $e^{-iHt}$ where, in the linear
regime, the Hamiltonian is given by $H=-\frac{1}{2}T_{00}h_{00}$. To the best of our knowledge, our derivation here is original, even though the result is well known. 

As a side remark, it is fruitful to think of the phase in analogy with the Aharonov Bohm phase, where the role of the $B$ field would be played here by $\Omega$. This is perfectly intuitive since we can think of the magnetic field as a form of curvature, just like the curvature that is responsible for the gravitational field in General Relativity. The metric in this case plays the role of the electromagnetic vector potential. 

So far we have only described a single particle superposed a fixed distance away from the centre of the rotating disk. Imagine now that the particle is superposed across two distances, $r_1$ and $r_2$, and that at each it completes interference (by being in a clockwise and anti-clockwise superposition - see the figure). 

\begin{figure*}[htb]
\begin{center}
\includegraphics[width=\columnwidth]{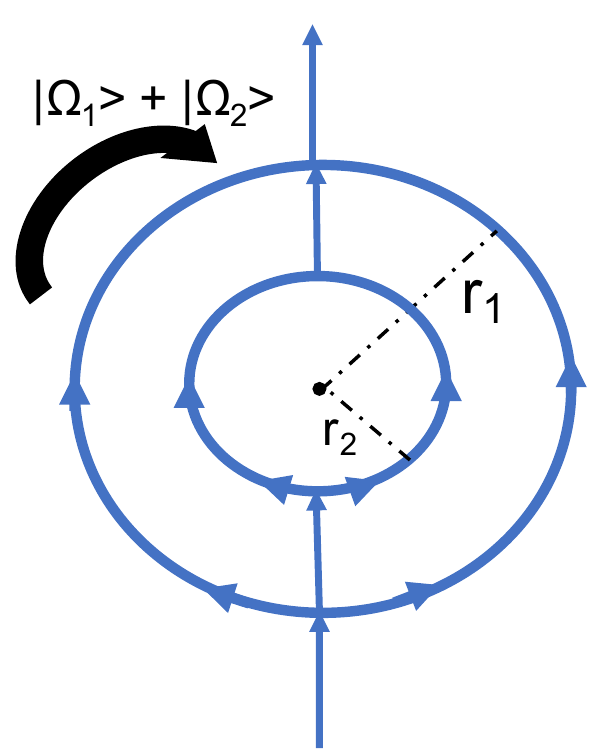}
\caption{Schematics of the quantum matter wave Sagnac interferometer with two different radii and two different angular frequencies $\Omega$. The same particle undergoing matter-wave interference is in a nested superposition of $r_1$ and $r_2$ and interfers at each radius before being recombined to yield the final state discussed in the text.}
\end{center}
\end{figure*}

Then at the end of the interference the state would be
\begin{equation}
e^{i\frac{2m}{\hbar} \times \Omega  \times A_1} |r_1\rangle + e^{i\frac{2m}{\hbar} \times \Omega \times A_2} |r_2\rangle \: .
\end{equation}
The detectable phase is then the difference between the two, i.e. $\frac{2m}{\hbar} \times \Omega_1 \times (A_2-A_1)$. However, imagine further that the disk is spinning in a state that is a
superposition of two frequencies $\Omega_1$ and $\Omega_2$ (just like an electron in an atom being in a superposition of two different angular momenta). Then the total state at the end of each complete interference is 
\begin{eqnarray}
e^{i\frac{2m}{\hbar} \times \Omega_1  \times A_1} |r_1\rangle |\Omega_1\rangle + e^{i\frac{2m}{\hbar} \times \Omega_2 \times A_1} |r_1\rangle |\Omega_2\rangle +\\ 
e^{i\frac{2m}{\hbar} \times \Omega_1  \times A_2} |r_2\rangle |\Omega_1\rangle + e^{i\frac{2m}{\hbar} \times \Omega_2 \times A_2} |r_2\rangle |\Omega_2\rangle 
\end{eqnarray}

In our notation, the state $|r\rangle|\Omega\rangle$ describes the radial and angular state of the particle. The state is initially clearly a product state of the two degrees of freedom (radial
and angular, being in a superposition of each), $(|r_1\rangle + |r_2\rangle ) (|\Omega_1\rangle + |\Omega_2\rangle )$. However, for a carefully chosen radii and angular frequencies, this state can after one circle become 
maximally entangled of the form  $|r_1\rangle (|\Omega_1\rangle + |\Omega_2\rangle ) + |r_2\rangle (|\Omega_1\rangle - |\Omega_2\rangle )$.

{\bf Discussion.}  In our previously proposed experiment \cite{MAVE}, the path entanglement between two interfering masses was proven to be a witness of the fact that the mediating gravitational field is non-classical. In the present experiment, both degrees of freedom that become entangled belong to the same particle. In this sense, it is at first hard to see what the mediator is. This becomes clear by considering that the spinning disk generates a curvature in space: each spinning frequency corresponds to a different curvature and therefore to a different gravitational field. Therefore, witnessing entanglement in this situation would, too, imply that the field has to be quantum, if one assumes that acceleration and gravitation are equivalent and that the equivalence principle satisfies linearity of quantum theory (meaning that it holds in each branch of the superposition). In order to explain the entanglement one has to resort to the superposition of two different spacetime geometries, one for each frequency, (similarly to what described in \cite{ROV}). This is an important example of the generality of our argument for the non-classicality of gravity as a mediator of entanglement, where it becomes clear that the locality assumption is to be intended more generally than in the context of a propagator between spatially separated locations (see also \cite{MAVE0}). 

A final thought-provoking remark: it might seem that the experiment involving a superposition of different rotations and different spatial locations is difficult to perform in practice. However, if we assume that the particle is an electron in say the Hydrogen atom, and that this electron is superposed across two different principal quantum numbers and across two different orbital quantum numbers, we have basically exactly the scenario we need. Then, $\delta \phi \approx m_e \Omega A/\hbar \approx 1$, and so we already have conditions in nature for the maximum entanglement to be created via the mechanism we suggested. If the equivalence principle holds in the form we have envisaged, then already some phases acquired in the ordinary atomic physics situation would be a proof of the quantum nature of gravity.

\textit{Acknowledgments}: This publication was made possible through the support of the ID 61466 grant from the John Templeton Foundation, as part of the The Quantum Information Structure of Spacetime (QISS) Project (qiss.fr). The opinions expressed in this publication are those of the authors and do not necessarily reflect the views of the John Templeton Foundation. CM thanks the Eutopia Foundation. VV thanks the National Research Foundation, Prime Minister's Office, Singapore, under its Competitive Research Programme (CRP Award No. NRF- CRP14-2014-02) and administered by Centre for Quantum Technologies, National University of Singapore.


\begin{thebibliography}{1}

%
\bibitem{SOUG} S. Bose, {\it et al.} Phys. Rev. Lett. 119, 240401 (2017).
%
\bibitem{MAVE}  C. Marletto and V. Vedral, Phys. Rev. Lett. 119, 240402 (2017).
%
\bibitem{BEI} C. Anastopoulos and B. L. Hu, Class. Quant. Grav. 35, 035011 (2018).
\bibitem{HARDY} L. Hardy, arXiv:1903.01289, (2019). 
\bibitem{BRU1} M. Zych, C. Brukner, Nat. Phys., {\bf 14}, 1027-1031, (2018).
\bibitem{BRU2} Rosi, G. et al., Nat. Commun. {\bf 8}, 15529 (2017).
\bibitem{SAGNAC} G. Sagnac, Comptes Rendus, {\bf 157}, 708  (1913).
%
\bibitem{MAVE0} C. Marletto and V Vedral, {\bf 3}, 29, npj Quantum Information, 2017.
%
\bibitem{MAVE1} C. Marletto and V. Vedral, Phys. Rev. D {\bf 98}, 046001 (2018).
%
\bibitem{ROV} M. Christodoulou and C. Rovelli, Phys. Rev. B, 792, 64-68, (2019.)
%

\end{thebibliography}
\end{document}